\documentclass[twocolumn]{jpsj3}
\newcommand{\lsim} 
 {\ \raise.35ex\hbox{$<$}\kern-0.75em\lower.5ex\hbox{$\sim$}\ }
\newcommand{\gsim}
 {\ \raise.35ex\hbox{$>$}\kern-0.75em\lower.5ex\hbox{$\sim$}\ }
\def\journal #1#2#3#4{#1 {\bf #2}, #3 (#4)}
\def\PRB{Phys.\ Rev.\ B}
\def\PRL{Phys.\ Rev.\ Lett.}

\def\SSC{Solid State Commun.}

\def\JPCS{J.\ Phys.\ Chem.\ Solids}

\def\JPSJ{J.\ Phys.\ Soc.\ Jpn.}

\def\SSC{Sol.\ St.\ Comm.}

\usepackage{graphicx}
\usepackage{dcolumn}
\usepackage{bm}
\usepackage{amsmath}
\usepackage{times}
\usepackage[usenames]{color}

\title{
Quantum Dielectric Fluctuation in an Electronic Ferroelectricity \\
studied by Variational Monte-Carlo Method
}
\author{ Tsutomu~Watanabe,$^{1 \ast}$ and Sumio~Ishihara$^{2,3}$ }
\inst{
$^1$Department of Natural Science, 
Chiba Institute of Technology, Narashino 275-0023, Japan \\
$^2$Department of Physics, Tohoku University, Sendai 980-8578, Japan \\
$^3$CREST, JST, Sendai 980-8578, Japan 
}
\date{\today}
\abst{
Electronic structure and dielectric property in an electronic ferroelectricty, where electric polarization is driven by an electronic charge order without inversion symmetry, are studied. 
Motivated from layered iron oxides, roles of quantum fluctuation on ferroelectricity in a paired-triangular lattice are focused on. 
Three types of the extended $V-t$ model are examined by the variational Monte-Carlo method with the Gutzwiller-type correlation factor. 
It is shown that electron transfer between the triangular layers corresponding to the inter-layer polarization fluctuation promotes the three-fold charge order associated with an electric polarization. 
This result is in highly contrast to the usual manner of quantum fluctuation in the hydrogen-bond type ferroelectricities and the quantum paraelectric oxides. 
Spin degree of freedom of electron and a realistic crystal structure for the layered iron oxides further stabilize the polar charge ordered state. 
Implications of the numerical results for layered iron oxides are discussed.  
}

\kword{Ferroelectricity, Frustration, Charge Order}

\begin{document}
\maketitle

%   Abstract
%-----------------------------------------------------------

%-----------------------------------------------------------

\section{Introduction\label{sec:intro}}
%%%%%%%%%%%%%%%%%%%%%%%%%%%%%%%%%%%%%%%%%%

Roles of electronic degree of freedom in ferroelectricity and related phenomena have attracted much attention for a long time. 
In particular, covalency contribution on the displasive-type ferroelectricity has been revealed by the modern $ab$-initio electronic-structure calculation and the Berry phase theory for the electric polarization. 
Recently discovered multiferroics, i.e. coexistence of ferroelectricty and magnetism, are another example.~\cite{Kimura,kobayashi03, hur04,Katsura,sergienko06, mostovoy06}
Through the intensive theoretical and experimental examinations, it has been uncovered that the symmetric/anti-symmetric exchange interactions and the exchange striction effect are origin of this type ferroelectricity. 

It is known that there is different type of ferroelectricity driven by electronic degree of freedom. 
A long-range order of the electronic charge without inversion symmetry is responsible for a macroscopic electric polarization. This phenomenon is termed the electronic ferroelectricity or the charge-order type ferroelectricity.~\cite{ishihara,khomskii}
A possibility of this type of ferroelectricity has been suggested experimentally in a variety of transition-metal oxides, e.g. Pr(Sr$_x$Ca$_{1-x}$)Mn$_2$O$_7$,~\cite{Tokunaga} LuFe$_2$O$_4$,~\cite{Ikeda2,Ikeda1} low-dimensional organic salts, e.g. TMTTF$_2$X ($X$=PF$_6$, AsF$_6$),~\cite{Monceau} $\alpha$-(BEDT-TTF)$_2$I$_3$~\cite{Yamamoto}, $\kappa$-(BEDT-TTF)$_2$Cu$_2$(CN)$_3$~\cite{sasaki,naka_kappa,hotta_kappa} and others. 
Most of this type of material belongs to the so-called quarter-filled system where the number of carrier per site is 0.5.  
The electronic ferroelectricity has some similarities to the hydrogen-bond type ferroelectricity; localization and delocalization of charged particles, i.e. electron or proton, are concerned in the ferroelectric transition. 
One noticeable difference between the two is that the electron mass is much smaller than the proton mass. Therefore, large dielectric fluctuation is expected to play crucial roles in the electronic ferroelectricity. 

Layered iron oxide $R$Fe$_2$O$_4$ ($R$=Lu, Yb, Y) is recognized as a representative electronic ferroelectricity. 
The crystal structure in $R$Fe$_2$O$_4$ consists of a stacking of the $R$-O layers and the Fe-O layers where Fe ions form the paired-triangular lattices.~\cite{Kimizuka} 
A nominal valence of an Fe ion is 2.5+, and an equal amount 
of Fe$^{2+}$ with a $d^6 (S=2)$ configuration and Fe$^{3+}$ with $d^5 (S=5/2)$ coexists.  
A charge order of Fe $3d$ electrons in LuFe$_2$O$_4$ was found by the x-ray and electron diffraction experiments \cite{Yamada1,Yamada2,Angst,Mulders}
where the superlattice peaks appear at $(n/3\ n/3\ 3m+1/2)$ below about 320K. 
A ferrimagnetic long-range order was confirmed by the neutron diffraction experiments where the magnetic peaks at $(1/3\ 1/3\ m)$ are observed below about 250K \cite{Akimitsu,Shiratori,Wu}. 
The temperature dependence of the electric polarization was measured by the pyroelectric current. 
It was found that the polarization appears around 320K and increases around 250K.~\cite{Ikeda1,Ikeda2} 
These results are interpreted that the electric polarization is deduced by the Fe$^{2+}$/Fe$^{3+}$ charge order, and 
is strongly coupled with the spins in Fe ions. 

%**********************************************************************
%\b[t]
%\vspace{-0.2cm}
%\begin{center}
%\includegraphics[width=1\columnwidth,clip]{polc}
%\includegraphics[width=4.0cm,height=1.2cm]{polc}
%\end{center}
%\vspace{-0.4cm}
%\caption{(Color online)
%Charge structure of the three-fold with electric dipole moments 
%presented in $R$Fe$_2$O$_4$. 
%Black and white circles indicate Fe$^{3+}$ and Fe$^{2+}$ ions, 
%respectively. 
%}
%\label{fig:polc}
%\vspace{-0.5cm}
%\end{figure}
%**********************************************************************

A primitive model for the observed charge order in $R$Fe$_2$O$_4$ associated with the electric polarization was first proposed by Yamada and coworkers \cite{Yamada1,Yamada2}. 
It was considered that Fe ions are aligned as $\cdots$ Fe$^{3+}$-Fe$^{3+}$- Fe$^{2+}$ $\cdots$ along one of the crystal axes in a upper triangular layer and as
$\cdots$ Fe$^{3+}$-Fe$^{2+}$-Fe$^{2+}$ $\cdots$ in a lower layer. 
As a result, charge imbalance in a paired triangular lattice induces the electric dipole moment. 
Recently, detailed microscopic calculations for stability of the polar CO structure have been presented.~\cite{nagano07,Nagano,Naka,Nasu,Xiang1,Xiang2}.
One of the present authors and coworkers presented a microscpic theory for the electronic structure and dielectric properties.~\cite{Nagano,Naka} 
% in $R$Fe$_2$O$_4$. 
They treated the electronic charge and spin degrees of freedom as classical variables in the calculation, and showed that the polar charge order does not appear at zero temperature and is stabilized by the thermal fluctuation effect. 
This result is consistent with the recent optical experiments where large charge fluctuation remains far below the charge ordering temperature \cite{Xu}. 

As well as the thermal fluctuation, the quantum fluctuation of the electronic charge, i.e. the electron transfer between the sites where the electronic potential takes its minima, is expected to have major contribution to this-type of ferroelectric transition. 
This is because most of this type of ferroelectricity is located at vicinity of the metal-insulator transition. 
Role of the quantum fluctuation on ferroelectricity has been examined for a long time in the hydrogen-bond type ferroelectricity \cite{Blinc,Gennes}, i.e. the proton tunneling between the double-potential well, and the quantum paraelectric oxides, such as SrTiO$_3$ and KTaO$_3$.~\cite{qpara} 
It is known that the quantum motion of protons/ions prevents the system from the long-range ferroelectric order.  
More remarkable quantum fluctuation effect is expected in the electronic ferroelectricity, because the electron mass is much smaller than the ion/proton mass. 
It is highly nontrivial whether the conventional quantum fluctuation effects seen in the hydrogen-bond type ferroelectricy and the quantum paralectric oxides are naively applicable to the electronic ferroelectricity or not.

The present paper addresses the issue of the quantum fluctuation effect in the electronic ferroelectricity, in particular, the dielectric and magneto-dielectric properties in the layered iron oxides. 
We introduce the three-types of the extended $V-t$ model in a paired-triangular lattice and focus on the electron transfer effect between the layers, corresponding to the quantum fluctuation of the electric dipole moment. 
The models are analyzed by the variational Monte-Carlo (VMC) method by which the quantum fluctuation effect, the long-range Coulomb interaction and the geometrical frustration are treated properly. 
It is shown that the electron transfer between the layers tends to stabilize the polar charge order, in contrast to the conventional manner of the quantum fluctuation effect on the ferroelectricity. 
The spin degree of freedom and the realistic lattice structure in $R$Fe$_2$O$_4$ reinforce the electric polarization. 

In Sect. II, the extended $V-t$ model in the paired-triangular lattice and the formulation of the VMC method are introduced, and the numerical results in this model are presented. 
In Sect.~III, roles of the spin degree of freedom in the dielectric and magneto-dielectric properties are examined. 
In Sect.~IV, the numerical results in the model, where the realistic crystal structure in $R$Fe$_2$O$_4$ is  taken into account, are presented. 
Section V is devoted to summary and discussion. 
A part of the present results were briefly reported in Ref.~\cite{Watanabe}.

%%%%%%%%%%%%%%%%%%%%%%%%%%%%%%%%%%%%%%%%%%
\section{Electric Polarization in $V-t$ Model \label{sec:charge}}
%%%%%%%%%%%%%%%%%%%%%%%%%%%%%%%%%%%%%%%%%%
%%%%%%%%%%%%%%%%%%%%%%%%%%%%%%%%%%%%%%%%%%
%\section{Model and Method \label{sec:method}}
%%%%%%%%%%%%%%%%%%%%%%%%%%%%%%%%%%%%%%%%%%
 
\begin{figure}[t]
%\vspace{-0.2cm}
\begin{center}
\includegraphics[width=5.2cm,height=3.2cm]{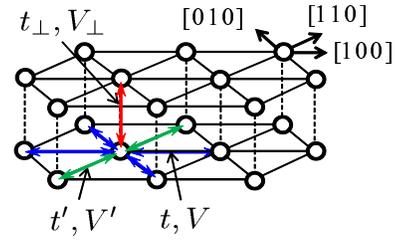}
\end{center}
\vspace{-0.4cm}
\caption{(Color online)
A pair of the triangular lattices 
and interactions in ${\cal H}_{Vt}$. 
}
\label{fig:mod1}
%\vspace{-0.5cm}
\end{figure}

As mentioned in the previous section, the crystal lattice of $R$Fe$_2$O$_4$ consists of the alternate stacking of the Fe-O double-triangle layers, termed W-layer, and the $R$-O layers along the $c$ axis. A unit cell with the R$\bar 3$c symmetry includes the three W-layers and the three $R$-O layers. 
We focus on electronic structure in a W-layer which dominates the electric and magnetic properties in $R$Fe$_2$O$_4$. We consider the three types of the $V-t$ model in the W-layer, 
and start from the spin-less $V-t$ model in a regularly stacked paired-triangular lattice. 
The double triangular lattices are stacked along the $c$ axis as shown in Fig.~\ref{fig:mod1}. 
The long-range Coulomb interactions and the inter-site electron transfers between the spin-less fermions are taken into account. 
A way of a stacking of the triangular lattices is somewhat different from that in $R$Fe$_2$O$_4$. 
Effects of the spin degrees of freedom and the realistic stacking of the triangular layers will be introduced in Sects.~\ref{sec:magnetic} and \ref{sec:real}, respectively. 

%**********************************************************************
%**********************************************************************

The model Hamiltonian is given as 
%----------------------------------------------------------------------
\begin{eqnarray}
{\cal H}_{Vt} = {\cal H}_t + {\cal H}_V, 
\label{eq:HVt} 
\end{eqnarray}
%----------------------------------------------------------------------
with 
%----------------------------------------------------------------------
\begin{eqnarray}
{\cal H}_t= 
&-&t  \sum\limits_{<ij>  m} c_{m i}^{\dag} c_{m j} 
-t' \sum\limits_{(ij)  m} {c_{m i}^{\dag} c_{m j} } \nonumber\\ 
&-&t_\bot \sum\limits_{i } c_{u i}^{\dag } c_{l i} +H.c. , 
\label{eq:Ht} 
\end{eqnarray}
%----------------------------------------------------------------------
and 
%----------------------------------------------------------------------
\begin{eqnarray}
{\cal H}_V &=& 
 V  \sum_{<ij> m} n_{m i} n_{m j}  
+V' \sum_{(ij) m} n_{m i} n_{m j} \nonumber \\ 
&+&V_\bot \sum\limits_i n_{u i} n_{l i}, 
\label{eq:Hv} 
\end{eqnarray}
%----------------------------------------------------------------------
where 
$c^\dag_{m i}$ is the creation operator for 
a spin-less fermion at site $i$ on the upper layer $(m=u)$ 
or the lower layer $(m=l)$, 
and $n_{m i}= c_{m i}^\dag c_{m i}$ is the number operator. 
A subscript $i$ takes a two-dimensional coordinate in a triangular lattice. 
To examine roles of frustration, 
we consider an anisotropic triangular lattice where bonds along the [110] direction are inequivalent to the bonds along [100] and [010] as shown in Fig.~\ref{fig:mod1}. 
We introduce the three kinds of the transfer integrals $t$, $t'$ and $t_\bot$ for the [100]/[010], [110] and [001] directions, respectively. 
In the same way, 
we introduce the three kinds of the inter-site Coulomb interactions $V$, $V'$ and $V_\bot$ for the nearest-neighbor (NN) pairs. 
Symbols $\sum_{\left\langle {ij} \right\rangle}$ and 
$\sum_{\left( {ij} \right)}$ 
represent summations for the NN pairs in the [100]/[010] and [110] directions, respectively. 
Ratios $V'/V$ and $t'/t$ represent magnitude of frustration. 
The number of the fermion per site is 0.5. 

We analyze this model by using the VMC method.  
A variational wave function is given by a following product form, 
\begin{equation}
\Psi={\cal P}\Phi 
\label{eq:tri}, 
\end{equation}
where $\Phi$ is the one-body Hartree-Fock (HF) part and ${\cal P}$
is the many-body correlation factor. 
We introduce the Gutzwiller-type correlation factor given by  
%----------------------------------------------------------------------
%\begin{eqnarray}
%{\cal P}\equiv \prod\limits_{mm'ij} 
%{\left ({1-v_{ij} n_{mi} n_{m'j}} \right)}, 
%\label{eq:Pv} 
%\end{eqnarray}
%----------------------------------------------------------------------
%----------------------------------------------------------------------
\begin{align}
{\cal P} &=
\prod\limits_{m <ij>} {\left ({1-v  n_{mi} n_{mj}} \right)} 
\prod\limits_{m (ij)} {\left ({1-v' n_{mi} n_{mj}} \right)}
\nonumber \\
& \times 
\prod\limits_{i} {\left ({1-v_\bot n_{ui} n_{li}} \right)}, 
\label{eq:Pv} 
\end{align}
%----------------------------------------------------------------------
%where $v_{ij}$ takes one of the three variational parameters 
%to be optimized, $v$, $v'$ and $v_\bot$, 
%when the NN $ij$ pairs are along the [100] and [010] directions, 
%the [110] one and the [001] one, respectively. 
where $v$, $v'$ and $v_\bot$ are the variational parameters. 
%
%**********************************************************************
\begin{figure}[t]
%\vspace{-0.2cm}
\begin{center}
\includegraphics[width=7.5cm,height=6.0cm]{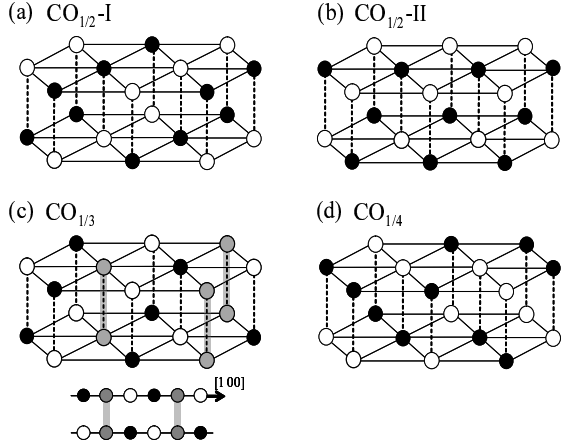}
\end{center}
\vspace{-0.4cm}
\caption{(Color online)
Schematic CO structures cosidered in ${\cal H}_{Vt}$: 
(a) CO$_{1/2}$-I, (b) CO$_{1/2}$-II, 
(c) CO$_{1/3}$ and (d) CO$_{1/4}$. 
The inset of (c) is a side view along the [120] direction. 
%\rd{((b) and (d) are the same.)}
}
\label{fig:CO}
%\vspace{-0.5cm}
\end{figure}
%**********************************************************************
As shown in Fig.~\ref{fig:CO}, 
in the one-body part of the wave function, we consider the following four-types CO structures: 
(i) a two-fold CO along the [110] direction termed CO$_{1/2}$-I, 
(ii) a two-fold CO along [100] termed CO$_{1/2}$-II, 
(iii) a three-fold CO along [110] termed CO$_{1/3}$, and 
(iv) a four-fold CO along [110] termed CO$_{1/4}$. 
These CO structures are the candidates for the mean-field (MF) ground state. 
In the CO$_{1/2}$-I, CO$_{1/2}$-II and CO$_{1/4}$ structures, 
stackings of the CO patterns in the two triangular-layers are out-of-phase. 
In each CO, the one-body part of the wave function is given by the HF wave function where the corresponding CO's are assumed in the MF order parameter. 
We take that black, white and grey circles in Fig.~\ref{fig:CO} correspond to the MF charge densities of  
$\langle n_{m i} \rangle-1/2=\Delta$, $-\Delta$ and zero, respectively.  
The MF order parameter $\Delta$ is regarded as a variational parameter and is optimized in $\Phi$.  
For example, $\Phi$ for the CO$_{1/3}$ structure is obtained by 
diagonalizing the following HF Hamiltonian, 
%
%----------------------------------------------------------------------
\begin{eqnarray}
{\cal H}_{V-t}^{\rm 1/3} &=& 
\sum\limits_{{\bf k}m} {{\bf{\phi}}_{{\bf k}m}^\dag h_{{\bf k}m}} 
{\bf{\phi}}_{{\bf k}m}  \nonumber \\
&-& t_\bot \sum\limits_{{\bf{k}}\lambda} {(c_{u \lambda {\bf{k}}}^{\dag} 
c_{l \lambda {\bf{k}}} + c_{l \lambda {\bf{k}}}^{\dag} 
c_{u \lambda {\bf{k}}})}, 
\label{eq:H3} 
\end{eqnarray}
%----------------------------------------------------------------------
where
%----------------------------------------------------------------------
\begin{eqnarray}
{\bf{\phi}}_{{\bf{k}}m}  = \left( {\begin{array}{*{20}c}
   {c_{m A {\bf{k}}} }  \\
   {c_{m B {\bf{k}}} }  \\
   {c_{m C {\bf{k}}} }  \\
\end{array}} \right), 
\label{eq:H3t} 
\end{eqnarray}
%----------------------------------------------------------------------
%----------------------------------------------------------------------
\begin{eqnarray}
h_{{\bf{k}}u}  = \left( {\begin{array}{*{20}c}
   {W } & {T_{\bf{k}} } & {T_{\bf{k}}^ *  }  \\
   {T_{\bf{k}}^ *  } & { - W } & {T_{\bf{k}} }  \\
   {T_{\bf{k}} } & {T_{\bf{k}}^ *  } & 0  \\
\end{array}} \right), 
\label{eq:B} 
\end{eqnarray}
%----------------------------------------------------------------------
%----------------------------------------------------------------------
\begin{eqnarray}
h_{{\bf{k}}l}  = \left( {\begin{array}{*{20}c}
   { - W } & {T_{\bf{k}} } & {T_{\bf{k}}^ *  }  \\
   {T_{\bf{k}}^ *  } & {W } & {T_{\bf{k}} }  \\
   {T_{\bf{k}} } & {T_{\bf{k}}^ *  } & 0  \\
\end{array}} \right). 
\label{eq:B2} 
\end{eqnarray}
%----------------------------------------------------------------------
We define 
%----------------------------------------------------------------------
\begin{eqnarray}
T_{\bf{k}} = - t(e^{ik_1} + e^{ik_2}) - t'e^{-i(k_1 + k_2)}  , 
\label{eq:A}
\end{eqnarray}
%----------------------------------------------------------------------
and 
%----------------------------------------------------------------------
\begin{eqnarray}
W = - \Delta (2V + V' + V_\bot) ,  
\label{eq:VMF}
\end{eqnarray}
%----------------------------------------------------------------------
where $k_1$ and $k_2$ are the wave-vectors along the $[100]$ and $[010]$ directions, respectively. 
A subscript $\lambda(={\rm A, B, C})$ represents the three sublattices in the CO$_{1/3}$ structure shown in Fig.~\ref{fig:CO}(c), and 
the summation $\sum\nolimits_{\bf k}$ runs over the reduced two-dimensional Brillouin zone. 
The operator $c_{m \lambda {\bf{k}}}$ is the Fourier transform of $c_{m \lambda i}$ where site coordinate is redefined by the two subscripts $\lambda$ and $i$. 
The one-body parts of the wave-functions for other CO's are defined in the same way.  
A polar CO$_{1/3}$ structure of the present interest has a possibility to be realized in the CO$_{1/3}$ structure, when all the grey circles in the upper (lower) plane in Fig.~\ref{fig:CO}(c) becomes charge rich (poor) sites due to the correlation factor ${\cal P}$ in a case of the full electric polarization. 

In the numerical calculation, we adopt $2\times 10^{5}-5\times$10$^5$ samples in most of the VMC simulations.  
The fixed-sampling method is used to optimize the variational parameters \cite{Umrigar}. 
An error of the energy expectation value is of the order of $10^{-4}t$. 
Cluster sizes are of $N_s = L\times L\times 2 (\equiv 2N)$ $(L\le 12)$ 
sites with the periodic-boundary condition and the antiperiodic-boundary condition. 
%The number of particles is fixed to be $N_e=N_s/2$. 

%%%%%%%%%%%%%%%%%%%%%%%%%%%%%%%%%%%%%%%%%%
%\section{Simple spin-less Fermion model \label{sec:charge}}
%%%%%%%%%%%%%%%%%%%%%%%%%%%%%%%%%%%%%%%%%%

%**********************************************************************
\begin{figure}[t]
%\vspace{-0.2cm}
\begin{center}
\includegraphics[width=7.0cm,height=5.0cm]{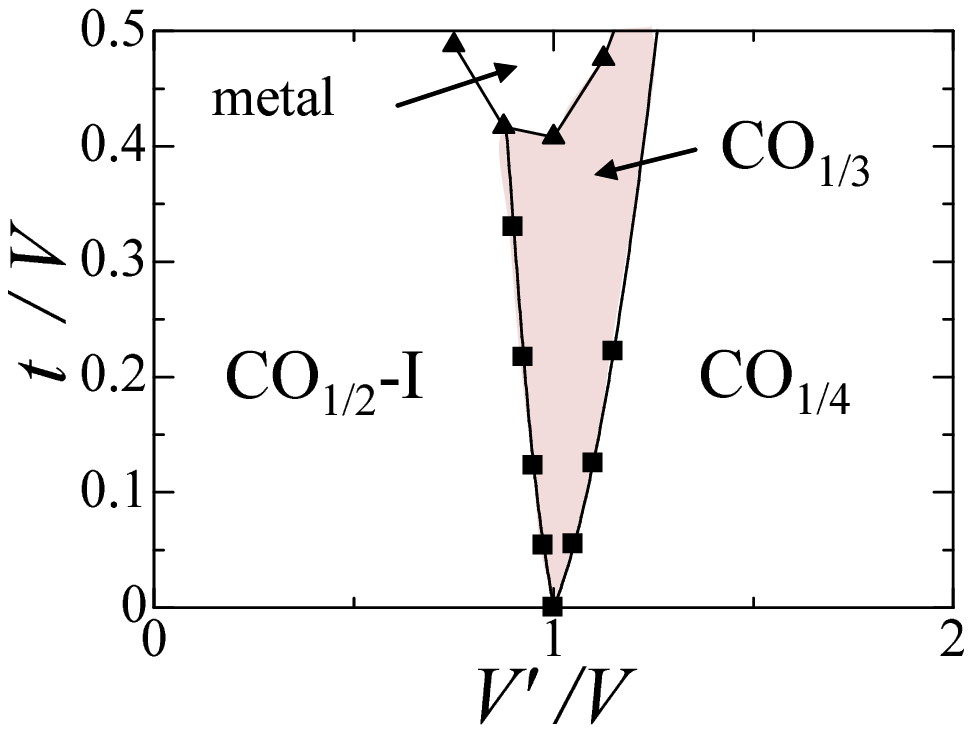}
\end{center}
\vspace{-0.5cm}
\caption{(Color online)
Phase diagram in ${\cal H}_{Vt}$. 
%in the plane of $t/V$ and $V'/V$. 
The relations $t=t'=t_\bot$ and $V = V_\bot$ are imposed. 
%\rd{Calculating the data up to $t/V=0.5$.}
}
\label{fig:simph}
%\vspace{-0.5cm}
\end{figure}
%**********************************************************************

In Fig.~\ref{fig:simph}, the phase diagram is presented 
in the plane of $V'/V$ and $t/V$. 
We chose $t=t'=t_\bot$ and $V=V_\bot$. 
This phase diagram is obtained from the calculated energy and charge correlation function defined by 
%----------------------------------------------------------------------
\begin{eqnarray}
N({\bf{q}}) = \frac{1}{{N_{\rm{s}} }} \sum\limits_{ijm} 
{e^{i{\bf{q}} \cdot ({\bf{R}}_{mi} - {\bf{R}}_{mj} ) }
\left ( \left\langle n_{mi} n_{mj}  \right\rangle - n^2 \right)}, 
\label{eq:Nq} 
\end{eqnarray}
%----------------------------------------------------------------------
where $n=1/2$ and ${\bf R}_{mi}$ is a position of site $i$ on the $m$ layer. 
In the classical limit, i.e. $t=0$, the CO$_{1/2}$-I and CO$_{1/4}$ structures are realized in the regions of $V'/V<1$ and $V'/V>1$, respectively. 
At $V'/V=1$ with $t=0$, three CO structures, i.e. 
CO$_{1/2}$-II, CO$_{1/4}$, and 
CO$_{1/3}$, are degenerate. 
When the electron transfer $t$ is introduced, 
the CO$_{1/3}$ structure is stabilized in a finite parameter region.

%**********************************************************************
\begin{figure}[t]
%\vspace{-0.2cm}
\begin{center}
\includegraphics[width=7.5cm,height=10.0cm]{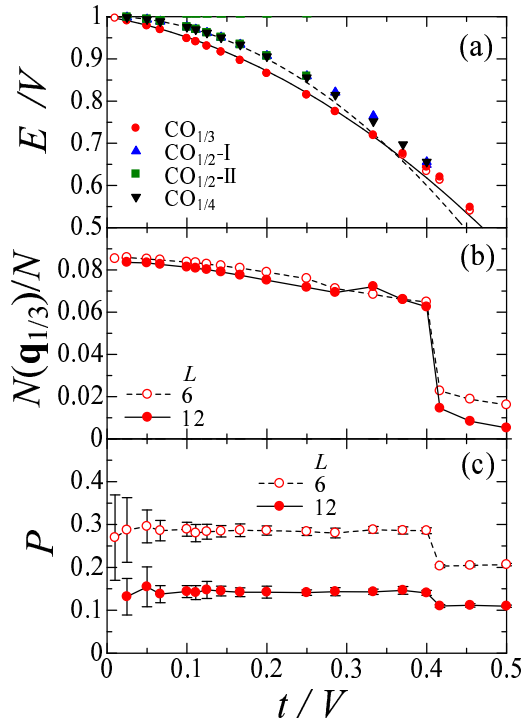}
\end{center}
\vspace{-0.5cm}
\caption{(Color online)
(a) Energies for several CO's. 
Solid and broken lines are for the energies obtained by the second-order perturbation in CO$_{1/3}$ and other CO's, respectively. 
(b) Charge correlation function $N({\bf q})$ at 
${\bf q}=(2\pi/3, 2\pi/3)$. 
(c) Polarization correlation $P$. 
Relations $t=t'=t_\bot$ and $V = V'= V_\bot$ are imposed.  
Open and filled symbols are for the results in $L=6$ and $12$, respectively. 
}
\label{fig:enqp}
%\vspace{-0.5cm}
\end{figure}
%**********************************************************************
%
%**********************************************************************
\begin{figure}[t]
%\vspace{-0.2cm}
\begin{center}
\includegraphics[width=7.5cm,height=7.0cm]{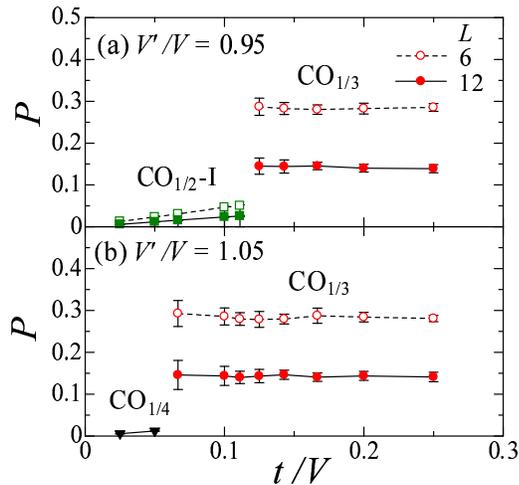}
\end{center}
\vspace{-0.5cm}
\caption{(Color online)
Polarization correlation $P$. 
Parameters are chosed to be $V'/V=0.95$ in (a) and $1.05$ in (b). 
Relations $t=t'=t_\bot$ and $V = V_\bot$ 
are imposed. 
Open and filled symbols are for the results in $L=6$ and $12$, respectively. 
}
\label{fig:difP}
%\vspace{-0.5cm}
\end{figure}
%**********************************************************************
%
Let us focus on the results in $V'/V=1$ in more detail. 
In Fig.~\ref{fig:enqp}(a), the energy expectation 
$E \equiv N^{-1} \langle {\cal H}_{Vt} \rangle$ is plotted as 
a function of $t/V$. 
Reduction of $E$ in CO$_{1/3}$ with increasing $t/V$ is more remarkable than those in the other CO structures. 
A stabilization of the CO$_{1/3}$ structure is attributed to the $t$-linear dependence of $E$, in contrast to the quadratic dependence in other CO's. 
The charge correlation function at 
${\bf q}=(2\pi/3, 2\pi/3) (\equiv {\bf q}_{1/3})$ is presented in Fig.~\ref{fig:enqp}(b). 
A possible maximum value of $N({\bf q}_{1/3})$ is $1/9N$ for the CO$_{1/3}$ structure. 
Magnitude of $N({\bf q}_{1/3})$ is about $70\%$ of its maximum value below $t/V=0.4$ and is almost independent of the system size $L$. 
This correlation almost disappears above $t/V=0.4$ where the metallic phase appears as shown later.  
In Fig.~\ref{fig:enqp}(c), 
we plot the correlation function of the electric polarization defined by 
%----------------------------------------------------------------------
\begin{eqnarray}
P = \left\langle p^2 \right\rangle^{\frac{1}{2}}, 
\label{eq:Pl} 
\end{eqnarray}
%----------------------------------------------------------------------
with 
%----------------------------------------------------------------------
\begin{eqnarray}
p = \frac{3}{N}
\sum_i
\left( n_{ui} -  n_{li} \right). 
\label{eq:Ps} 
\end{eqnarray}
%----------------------------------------------------------------------
A possible maximum value of $P$ in CO$_{1/3}$ is one. 
We also present the results for $V'/V=0.95$ and $1.05$ in Fig.~\ref{fig:difP} 
to compare $P$'s in several CO's. 
Magnitude of $P$ in CO$_{1/3}$ is much larger than those in CO$_{1/2}$-I and CO$_{1/4}$.
However, values of $P$ decrease almost by half with increasing the system size $L$ from $6$ to $12$. These results indicate that the electric polarization in the thermodynamic limit is supposed to be much smaller than the maximum value. 
We conclude that a large polarization fluctuation appears in the CO$_{1/3}$ phase, in comparison with other CO's, but the robust electric polarization is not expected in the thermodynamic limit. 

As explained below, the energy gain in CO$_{1/3}$ is caused by the inter-layer electron transfer, $t_\bot$. 
Let us focus on a pair of sites represented by grey circles 
in the upper and lower planes in Fig.~\ref{fig:CO}(c).  
Since these sites are surrounded by the in-plane three NN charge-rich sites and the NN three poor sites, the Coulomb interactions between NN sites are canceled out, and the classical energy does not depend on the charge configurations at the pair of grey-circle sites. 
Thus, the stabilization of the CO$_{1/3}$ structure is caused by the first order of $t_{\bot}$ between the grey-circle sites. 
This situation does not realize in other CO structures. 
Even when the longer-range Coulomb interactions than 
$V$, $V'$ and $V_\bot$ are taken into account,  
the above scenario for stabilization of CO$_{1/3}$ is expected to survive in the case where $t_\bot$ is larger than those interactions. 
To support this scenario, the energy expectations are calculated in 
the second-order perturbation with respect to $t$ for several CO's. 
The results are shown in Fig.~\ref{fig:enqp}(a). 
In the region of small $t$, the $t$-linear dependence of $E$ in CO$_{1/3}$ obtained by the VMC method is well reproduced by the perturbational calculation. 
It is concluded that the CO$_{1/3}$ is stabilized by a combined effect of the 
the geometrical frustration effect in the triangular lattice and the electron transfer between the layers. 

\begin{figure}[t]
%\vspace{-0.2cm}
\begin{center}
\includegraphics[width=7.5cm,height=5.0cm]{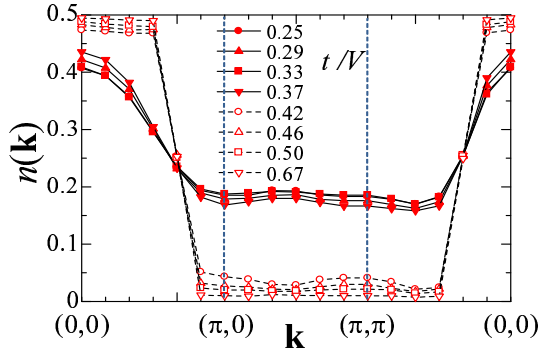}
\end{center}
\vspace{-0.5cm}
\caption{(Color online)
The momentum distribution function $n({\bf k})$ for various values of $t/V$. 
%along the path $(k_1,k_2)=(0,0)$-$(\pi,0)$-$(\pi,\pi)$-$(0,0)$. 
Relations $t=t'=t_\bot$ and $V = V' = V_\bot$ 
are imposed. System size is $L=12$. 
}
\label{fig:nk}
%\vspace{-0.5cm}
\end{figure}

We briefly touch the metal-insulator transition in this model. 
Discontinuous changes in $P$ and $N({\bf q}_{1/3})$ at $t/V \sim 0.4$ shown in Figs.~\ref{fig:enqp}(b) and 4(c) are associated with the metal-insulator transition. 
This is directly confirmed by calculating the momentum-distribution function 
defined by 
%----------------------------------------------------------------------
\begin{eqnarray}
n({\bf{k}}) = \frac{1}{{N_{\rm{s}} }}\sum\limits_{mij} {e^{i{\bf{k}} \cdot ({\bf{R}}_i  - {\bf{R}}_j )} \left\langle {c_{mi}^\dag  c_{mj} } \right\rangle }. 
\label{eq:nk} 
\end{eqnarray}
%----------------------------------------------------------------------
In Fig.~\ref{fig:nk}, $n({\bf k})$ at $V'/V=1$ is shown 
for various values of $t/V$ along the high-symmetry lines in the Brillouin zone. 
In $t/V>0.4$, discontinuous changes are seen in $n({\bf k})$ along $(0,0)$-$(\pi,0)$ and $(\pi,\pi)$-$(0,0)$.  

%**********************************************************************
%**********************************************************************

Let us compare the present results with the ones in the single-layer $V-t$ model.  
It was reported in Refs.~\cite{Hotta1,Hotta2,Miyazaki} that 
a long-range three-fold CO coexists with a metallic character in the case of $V=V'$ and $t=t'$ 
in the single-layer $V-t$ model. 
This is termed a "pinball liquid" (PBL) state. 
Here we set up the trial wave function for the PBL state in the paired-triangular lattice obtained by diagonalizing the following HF Hamiltonian, 
%--------------------------------------------------------------------
\begin{eqnarray}
{\cal H}_{V-t}^{P~1/3} &=& 
\sum\limits_{{\bf k}m} {{\bf{\phi}}_{{\bf k}m}^\dag h_{{\bf k}m}^{P}} 
{\bf{\phi}}_{{\bf k}m}  \nonumber\\
&-& t_\bot \sum\limits_{{\bf{k}}\lambda} {(c_{u \lambda {\bf{k}}}^{\dag} 
c_{l \lambda {\bf{k}}} + c_{l \lambda {\bf{k}}}^{\dag} 
c_{u \lambda {\bf{k}}})}, 
\label{eq:pliq} 
\end{eqnarray} 
%--------------------------------------------------------------------
where
%----------------------------------------------------------------------
\begin{eqnarray}
h_{{\bf{k}}u}^{P}  = \left( {\begin{array}{*{20}c}
   {W_1^P } & {T_{\bf{k}} } & {T_{\bf{k}}^ *  }  \\
   {T_{\bf{k}}^ *  } & {W_2^P} & {T_{\bf{k}} }  \\
   {T_{\bf{k}} } & {T_{\bf{k}}^ *  } & {W_3^P}  \\
\end{array}} \right), 
\label{eq:BP} 
\end{eqnarray}
%----------------------------------------------------------------------
and 
%----------------------------------------------------------------------
\begin{eqnarray}
h_{{\bf{k}}l}^{P}  = \left( {\begin{array}{*{20}c}
   {W_2^P} & {T_{\bf{k}} } & {T_{\bf{k}}^ *  }  \\
   {T_{\bf{k}}^ *  } & {W_1^P} & {T_{\bf{k}} }  \\
   {T_{\bf{k}} } & {T_{\bf{k}}^ *  } & {W_3^P}  \\
\end{array}} \right). 
\label{eq:BP2} 
\end{eqnarray}
%----------------------------------------------------------------------
We introduce $k_1$ and $k_2$ as the wave-vectors along the $[100]$ and $[010]$ directions, respectively, 
and define 
%----------------------------------------------------------------------
\begin{eqnarray}
W_1^P = - \gamma_p\Delta (2V + V' + \frac{1}{2}V_\bot),  
\label{eq:ple1}
\end{eqnarray}
%----------------------------------------------------------------------
%----------------------------------------------------------------------
\begin{eqnarray}
W_2^P =   \gamma_p\Delta (V + \frac{1}{2}V'+ V_\bot),  
\label{eq:ple2}
\end{eqnarray}
%----------------------------------------------------------------------
and 
%----------------------------------------------------------------------
\begin{eqnarray}
W_3^P =   \gamma_p\Delta (V + \frac{1}{2}V'- \frac{1}{2}V_\bot),  
\label{eq:ple3}
\end{eqnarray}
%----------------------------------------------------------------------
where $\gamma_p$ is a numerical factor taking $1$ and $-1$. 
When $\gamma_p=1$, the trial wave function represents the so-called electron-pinned PBL state where one of the three sublattices in the triangular lattice is occupied by 1/3-electrons per site and remaining 1/6-electrons per site move on other two subllatices. 
In this state, the mean fields in three sublattices 
are taken to be $\langle n_{u i} \rangle-1/2=(\Delta, -\Delta/2, -\Delta/2)$ and $\langle n_{l i} \rangle-1/2=(-\Delta/2, \Delta, -\Delta/2)$. 
When $\gamma_p=-1$, the wave function represents the so-called hole-pinned PBL state, 
where one of the three sublattices is occupied by 1/3-holes per site and 
1/6-holes per site move on others. 
The mean fields are taken to be 
$\langle n_{u i} \rangle-1/2=( -\Delta, \Delta/2, \Delta/2)$ and $\langle n_{l i} \rangle-1/2=(\Delta/2, -\Delta, \Delta/2)$. 

\begin{figure}[t]
%\vspace{-0.2cm}
\begin{center}
\includegraphics[width=7.5cm,height=5.0cm]{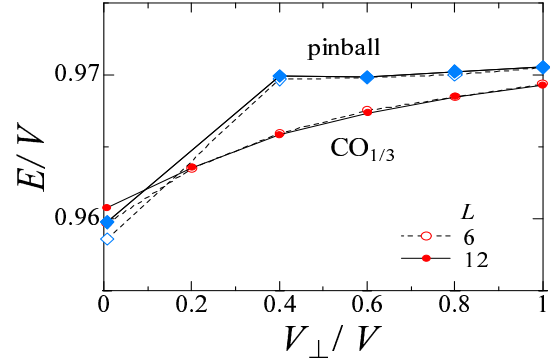}
\end{center}
\vspace{-0.5cm}
\caption{(Color online)
Energies in the hole-pinned pinball liquid state and the CO$_{1/3}$ state. 
Relations $t=t'=t_\bot$ and $V = V'$ are imposed.  
%\rd{In the trial pinball state, the mean-fields for the charge densities 
%are defined as $\langle n_{u i} \rangle-1/2=-\Delta$, 
%$\Delta/2$ and $\Delta/2$ 
%($\langle n_{l i} \rangle-1/2=\Delta/2$, $-\Delta$ and $\Delta/2$) 
%in the upper (lower) plane.(In the text.)} 
Open and filled symbols are for the results in $L=6$ and $12$, 
respectively. 
}
\label{fig:pl}
%\vspace{-0.5cm}
\end{figure}

The stabilities for the two PBL states are compared with the CO$_{1/3}$ state. 
The electron-pinned PBL has much higher energy in comparison with CO$_{1/3}$ 
in the present parameter range. 
In Fig.~\ref{fig:pl}, we present the energy expectations of the hole-pinned PBL state and the CO$_{1/3}$ state at $t=t'=t_\bot$ and $V=V'$. 
The PBL is realized only in a region of the small inter-layer Coulomb interaction, $V_\bot/V \lsim 0.1$, where the two triangular layer is almost independent. 
This is consistent with the previous results in the single triangular 
layer $V-t$ model analyzed by the VMC method.~\cite{Miyazaki}

%This is the so-called hole-pinned pinball state where one of the three sublattices in the single triangular lattice is occupied by 1/3-holes per site and 
%1/6-holes per site move on the rest of sublattices,
%The so-called electron-pinned pinball state where **** 
%has much higher energy in comparison with CO$_{1/3}$ in the present parameter range. 

%**********************************************************************
%**********************************************************************

%%%%%%%%%%%%%%%%%%%%%%%%%%%%%%%%%%%%%%%%%%
\section{Role of Spin Order 
\label{sec:magnetic}}
%%%%%%%%%%%%%%%%%%%%%%%%%%%%%%%%%%%%%%%%%%

So far, we examine the spin-less $V-t$ model in a paired triangular lattice. 
It is shown that the inter-plane transfer integral stabilizes the CO$_{1/3}$ phase where a larger polarization fluctuation appears in comparison with other types of CO's. 
However, magnitude of the electric polarization in the thermodynamic limit is supposed to be much smaller than the possible maximum value. 
We consider that additional small factors, which are not taken into account so far, may stabilize the electric polarization in the CO$_{1/3}$ phase. 
In this section, we investigate roles of the spin degree of freedom on the dielectric and magneto-dielectric properties in a paired-triangular lattice. 

\begin{figure}[t]
%\vspace{-0.2cm}
\begin{center}
\includegraphics[width=7.0cm,height=2.2cm]{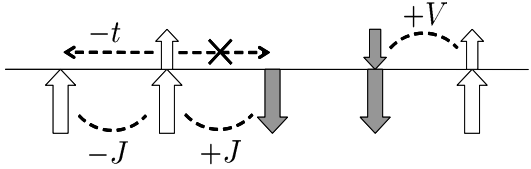}
\end{center}
\vspace{-0.5cm}
\caption{(Color online)
Interactions in ${\cal H}_{\rm S}$. 
Large and small arrows represent spin directions of the localized spins and the conduction 
electrons, respectively. 
}
\label{fig:smodel}
%\vspace{-0.5cm}
\end{figure}

\begin{figure}[t]
%\vspace{-0.2cm}
\begin{center}
\includegraphics[width=7.8cm,height=9.0cm]{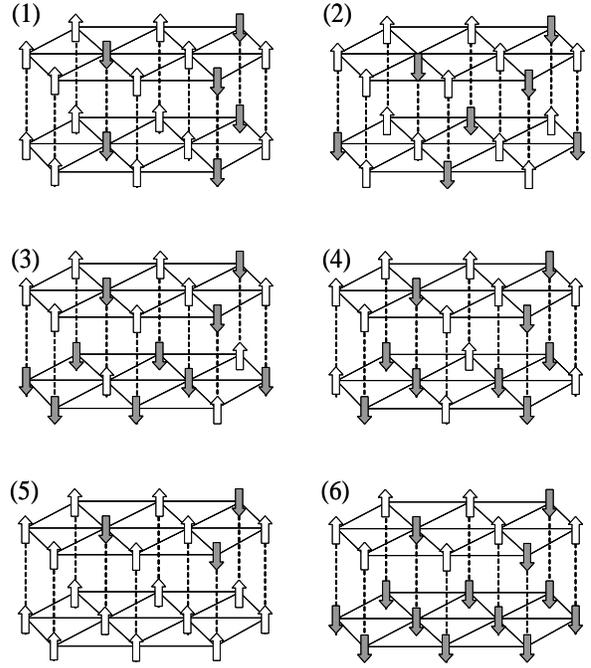}
\end{center}
\vspace{-0.4cm}
\caption{(Color online)
Configurations of the localized spin direction considered in ${\cal H}^{\rm S}$. 
}
\label{fig:sixp}
%\vspace{-0.5cm}
\end{figure}

We consider the spin degree of freedom on the basis of LuFe$_2$O$_4$ where two kinds of spin, $S=2$ for Fe$^{2+}$ and $S=5/2$ for Fe$^{3+}$, exist. 
The electron configurations of Fe$^{2+}$ and Fe$^{3+}$ are $3d^{6}$ and $3d^{5}$, respectively. It is thought that an excess electron in Fe$^{2+}$ occupy one of the doubly degenerate $d_{x^2-y^2}$ and $d_{xy}$ orbitals which have a larger transfer integral in the plane than other $3d$ orbitals.~\cite{Nagano,Naka}
We introduce, in the model Hamiltonian,  
a localized spin with $S=2$ at each site, itinerant carriers corresponding to the excess electrons in Fe$^{2+}$, 
and the ferromagnetic Hund couple between them at each site. 
The number of itinerant carriers per site is set to be 0.5.  
This is the generalized double-exchange model. 
For simplicity, we assume that the Hund coupling is infinite, and the localized spins are the Ising type denoted by $\sigma_{m i}$ $(m= u, l)$ which takes $\pm 1$. 
The Hamiltonian is given by 
%----------------------------------------------------------------------
\begin{eqnarray}
{\cal H}^{\rm S}={\cal H}_{t}^S+{\cal H}_{V}^S+{\cal H}_{\rm AFM} .
\label{eq:Hspin} 
\end{eqnarray}
%----------------------------------------------------------------------
The first and second terms represent the electron transfer and the inter-site Coulomb interaction, respectively,  given by 
%----------------------------------------------------------------------
\begin{eqnarray}
{\cal H}_{t}^S = 
&-& \sum\limits_{<ij>ms} {t_{\sigma _{mi} 
\sigma _{mj} } c_{mis}^\dag  c_{mjs} } - 
\sum\limits_{(ij)ms} {t'_{\sigma _{mi} \sigma _{mj} 
} c_{mis}^\dag  c_{mjs} } \nonumber\\
&-& \sum\limits_{is} {t_{ \bot \sigma _{mi} \sigma _{mj} } 
c_{uis}^\dag  c_{ljs} } + H.c., 
\label{eq:Hts} 
\end{eqnarray}
%----------------------------------------------------------------------
and 
%----------------------------------------------------------------------
\begin{eqnarray}
{\cal H}_{V}^S &=& 
 V  \sum\limits_{<ij>mss'} n_{mis} n_{mjs'}  
+V' \sum\limits_{(ij)mss'} n_{mis} n_{mjs'} \nonumber \\ 
&+&V_\bot \sum\limits_{iss'} n_{uis} n_{lis'}, 
\label{eq:Hvs} 
\end{eqnarray}
%----------------------------------------------------------------------
where $c_{m i s}$ is the electron annihilation operator at site $i$ on the $m$ layer 
with spin $s(=\uparrow, \downarrow)$.
% and $n_{mi}=c_{mi}^\dagger c_{mi} $ is the number operator.   
The third term in Eq.~(\ref{eq:Hspin}) is for the antiferromagnetic superexchange interaction between the localized spins given by 
%----------------------------------------------------------------------
\begin{eqnarray}
{\cal H}_{\rm AFM}
&=&J_{S} \biggl \{\sum\limits_{<ij>m}  \sigma_{mi} \sigma_{mj} 
 + \sum\limits_{(ij)m} \sigma_{mi} \sigma_{mj} \nonumber\\
&+&\sum\limits_i {\sigma_{ui} \sigma_{li}} \biggr \}, 
\label{eq:Haf} 
\end{eqnarray}
%----------------------------------------------------------------------
where the bond-dependence of the antiferromagnetic superexchange interaction $J_S$ is neglected. 
In ${\cal H}_{t}^S$, we consider that the transfer integrals depend on configuration of the localized spins. 
In the infinite-limit of the Hund coupling, 
it is reasonable to assume the following form of the transfer integral,  
\begin{eqnarray}
t_{\sigma_{mi} \sigma_{mj}}= 
\left \{ 
\begin{array}{ll}
t, & {\rm for}  \ \sigma_{mi}=\sigma_{mj} \\
0, &  {\rm for } \ \sigma_{mi} \ne \sigma_{mj} , 
\end{array} \right.
\end{eqnarray}
where $t$ is introduced in Eq.~(\ref{eq:Ht}). 
We also define $t'_{\sigma_{mi} \sigma_{mj}}$ and 
$t_{\bot \sigma_{mi} \sigma_{mj}}$ in ${\cal H}_t^S$ in the same way. 
A schematic view of the Hamiltonian is presented in Fig.~\ref{fig:smodel}. 
As for the localized spin structure, we consider all possible ordered structures characterized by the momenta $(2\pi/3, 2\pi/3)$ which corresponds to the observed neutron diffraction peak at $(1/3\ 1/3\ m)$ in LuFe$_2$O$_4$.~\cite{Akimitsu,Shiratori,Wu}
These magnetic structures are shown in Fig.~\ref{fig:sixp} 
where equivalent patterns are not shown. 

%**********************************************************************
%**********************************************************************

%**********************************************************************
%**********************************************************************

We analyze this model Hamiltonian by using the VMC method. 
The one-body part of the variational wave function $\Phi$ introduced in Eq.~(\ref{eq:tri}) for the CO$_{1/3}$ structure is obtained from the HF Hamiltonian. 
Here, we redefine the HF Hamiltonian in the three sublattices as follows, 
%----------------------------------------------------------------------
\begin{eqnarray}
{\cal H}^{\rm S \ 1/3} &= 
\sum\limits_{{\bf{k}}ms} {{\bf{\phi}}_{{\bf{k}}ms}^{\dagger} 
h_{{\bf k}m}^{\rm S} {\bf{\phi}}_{{\bf{k}}ms} } 
\nonumber\\ 
&- t_ \bot  \sum\limits_{{\bf{k}}\lambda s } 
{\eta_{\sigma_{u \lambda} \sigma_{l\lambda}}  
(c_{ u \lambda {\bf{k}}u }^{\dag } c_{l \lambda {\bf{k}}s} 
+ c_{l \lambda {\bf{k}}s}^{\dag } c_{u \lambda {\bf{k}}s} )}, 
\label{eq:Hmfs}
\end{eqnarray}
%----------------------------------------------------------------------
with 
%----------------------------------------------------------------------
\begin{eqnarray}
{\bf{\phi}}_{{\bf{k}}ms}  = \left( {\begin{array}{*{20}c}
   {c_{m A {\bf{k}} s}}  \\
   {c_{m B {\bf{k}} s} }  \\
   {c_{m C {\bf{k}} s} }  \\
\end{array}} \right), 
\label{eq:ts}
\end{eqnarray}
%----------------------------------------------------------------------
%----------------------------------------------------------------------
\begin{eqnarray}
h_{{\bf{k}}m}^{\rm S} =
\left( {\begin{array}{*{20}c}
   0 & {\gamma _{\sigma_{mA} \sigma_{mB}} T_{\bf{k}} } & {\gamma _{\sigma_{Cm}\sigma{Am}} T_{\bf{k}}^ *  }  \\
   {\gamma _{\sigma_{Am}\sigma_{Bm}} T_{\bf{k}}^ *  } & 0 & {\gamma_{\sigma_{Bm}\sigma_{Cm}} T_{\bf{k}} }  \\
   {\gamma_{\sigma_{Cm}\sigma_{Am}} T_{\bf{k}} } & {\gamma_{\sigma_{Bm}\sigma_{Cm}} T_{\bf{k}}^ *  } & 0  \\
\end{array}} \right), 
\label{eq:Bs}
\end{eqnarray}
%----------------------------------------------------------------------
%\rd{(Where is the V term?)}
where we introduce 
$\gamma_{\sigma_{\lambda m} \sigma_{\lambda' m}}=\delta_{\sigma_{m\lambda} 
\sigma_{m\lambda'} }$ and 
$\eta_{\sigma_{u \lambda} \sigma_{l \lambda}}= \delta_{\sigma_{u\lambda} \sigma_{l\lambda} }$. 
The correlation factor ${\cal P}$ is defined in Eq.~(\ref{eq:Pv}). 
The energy is optimized in each spin configuration shown in Fig.~\ref{fig:sixp}. A ratio of the number of the spin-up and spin-down electrons, $N_{\uparrow}/N_{\downarrow}$, is also optimized. 
%The localized and conduction electron spin pattern with the lowest energy 
%is regarded as the ground state. 
We restrict our calculations to the case of 
$V=V'=V_\bot$ and $t=t'=t_\bot$. 

%**********************************************************************
\begin{figure}[t]
%\vspace{-0.2cm}
\begin{center}
\includegraphics[width=7.5cm,height=5.5cm]{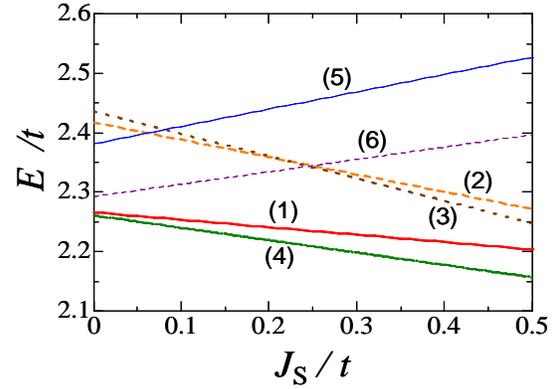}
\end{center}
\vspace{-0.5cm}
\caption{(Color online)
Energy optimized in each spin configuration in ${\cal H}^{\rm S}$. 
Numbers (1)-(6) correspond to the spin configurations given in Fig.~\ref{fig:sixp}. 
Parameters are chosen to be $t/V = 0.2$, $t = t'=t_\bot$ and $V = V'= V_\bot$. 
}
\label{fig:JP}
%\vspace{-0.5cm}
\end{figure}
%**********************************************************************

%**********************************************************************
\begin{figure}[t]
%\vspace{-0.2cm}
\begin{center}
\includegraphics[width=7.0cm,height=7.5cm]{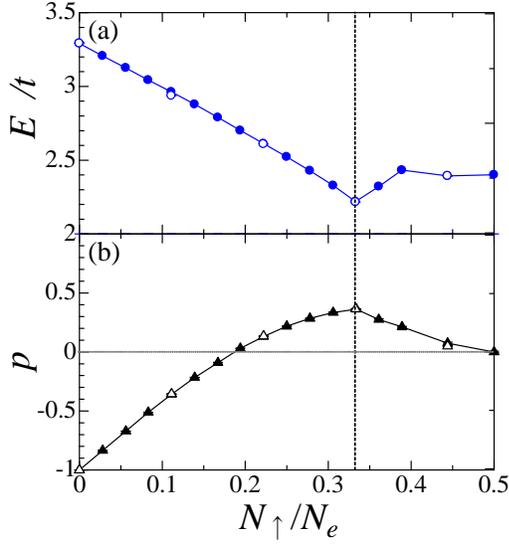}
\end{center}
\vspace{-0.5cm}
\caption{(Color online)
(a) Energy and (b) polarization correlation as functions of 
the number of the up-spin conduction electrons in ${\cal H}^S$. 
 Parameters are chosen to be $t/V = 0.2$, $J_{\rm S}/t = 0.05$, $t = t'=t_\bot$ and 
$V = V'= V_\bot$. 
Open and filled symbols are for the results in $L = 6$ and $12$, respectively. 
}
\label{fig:spinE}
%\vspace{-0.5cm}
\end{figure}
%**********************************************************************

In Fig.~\ref{fig:JP}, the optimized energies in the six spin configurations are plotted as a function of $J_{\rm S}$.
It is shown that the energy in the spin configuration 4 (see Fig.~\ref{fig:sixp}) is the lowest. 
In Fig.~\ref{fig:spinE}, 
the energy and the electric polarization correlation in the spin configuration (4) are plotted as a function of $N_\uparrow/N_e$ where $N_e=N_\uparrow+N_\downarrow$. 
At $N_\uparrow/N_e=0.33$, the energy minima and a large magnitude of the polarization are realized. 
These magnitudes are almost independence of the system size $L$. 
That is, the spin degree of freedom induces the robust electric polarization. 

%**********************************************************************
\begin{figure}[t]
%\vspace{-0.2cm}
\begin{center}
\includegraphics[width=8.0cm,height=4.0cm]{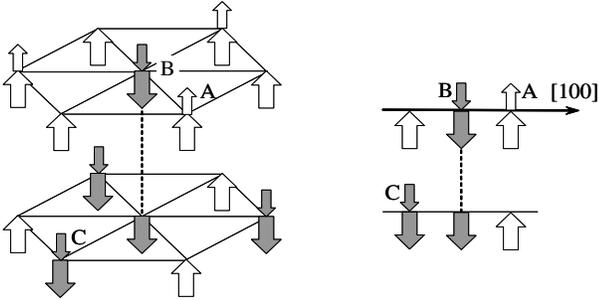}
\end{center}
\vspace{-0.4cm}
\caption{(Color online)
Left: spin structure in the optimized state corresponding to the energy minimum at $N_\uparrow/N_e=0.33$ 
in Fig.~\ref{fig:spinE}(a). 
% $t/V = 0.2$ and $J_{\rm S}/t = 0.05$. 
Large and small arrows represent spin directions of the localized spin and the conduction electrons, respectively. 
Right: a side view along the [120] direction. 
}
\label{fig:pmod}
%\vspace{-0.5cm}
\end{figure}
%**********************************************************************

The spin structure realized at the energy minima shown in Fig.~\ref{fig:spinE}(a) is presented in Fig.~\ref{fig:pmod}. 
Let us focus on sites A and B in the upper layer and consider the electron transfer from these sites to the sites just below them. 
When the spin degree of freedom is neglected, the classical energy is not changed by these electron transfer processes. 
This is the similar situation to the CO$_{1/3}$ structure in the spin-less $V-t$ model as explained in the previous section. 
Therefore, the energy in this structure is gained by the linear order of the electron transfer between the layers. 
However, in contrast to the previous spin-less case, the spin order suppresses the electron transfer between the layers and promotes the charge imbalance between the two layers as follows;  
1) In the upper layer, 2/3 of the localized spins are polarized toward the upper direction. Therefore, the up-spin conduction electrons (A) are confined in the upper layer to gain the in-plane kinetic energy. 
2) The transfer of the down-spin electron (B) from the upper layer to the lower layer is prevented, because increasing of the number of the down-spin electron in the lower layer enhances the kinetic energy. 
As the results, 2/3 of the electrons are confined in the upper layers and the charge imbalance between the layers occurs. 
That is, the electric and magnetic polarizations are induced cooperatively. 

%Here, it is worth noting that the optimized localized-spin configuration is invariant under the space and time inversions plus the in-plane translation along [100]. 
%Thus, the spontaneous spin- and charge-polarizations in the conduction electrons are not enforced by the broken symmetry of the localized-spin order. 

%%%%%%%%%%%%%%%%%%%%%%%%%%%%%%%%%%%%%%%%%%
\section{Effect of Realistic Crystal Structure in $R$Fe$_2$O$_4$
\label{sec:real}}
%%%%%%%%%%%%%%%%%%%%%%%%%%%%%%%%%%%%%%%%%%

\begin{figure}[t]
%\vspace{-0.2cm}
\begin{center}
\includegraphics[width=6.0cm,height=3.5cm]{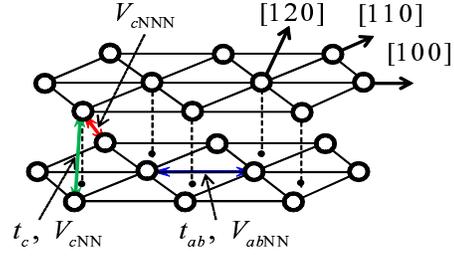}
\end{center}
\vspace{-0.5cm}
\caption{(Color online)
Lattice structure and interactions in ${\cal H}_{Vt}^{\rm R}$. 
In $R$Fe$_2$O$_4$, positions of the oxygen ions in the lower-layer are indicated by small black circles.
}
\label{fig:mod2}
%\vspace{-0.5cm}
\end{figure}

In this section, we introduce the realistic crystal structure in $R$Fe$_2$O$_4$ 
as a candidate to stabilize the electric polarization in the CO$_{1/3}$ structure.  
In $R$Fe$_2$O$_4$, a position of a Fe ion in the upper layer is not upon 
a Fe ion in the lower layer, but is upon an O ion surrounded 
by three in-plane Fe ions as shown in Fig.~\ref{fig:mod2}. 
We modify the Hamiltonian ${\cal H}_{Vt}$ defined in Eq.~(\ref{eq:HVt}) 
by taking this crystal structure into account. The model Hamiltonian is given by 

%**********************************************************************
%**********************************************************************
%
%----------------------------------------------------------------------
\begin{eqnarray}
{\cal H}^{\rm R} = {\cal H}_{t}^{\rm R} + {\cal H}_{V}^{\rm R}, 
\label{eq:HVtR} 
\end{eqnarray}
%----------------------------------------------------------------------
with 
%----------------------------------------------------------------------
\begin{eqnarray}
{\cal H}_t^{\rm{R}} =
&-&t_{{\rm{ab}}} \sum\limits_{ <ij> m} {c_{mi}^\dag c_{mj} }
 - t_{\rm{c}} \sum\limits_{(ij)} {c_{ui}^\dag c_{lj} } \nonumber\\
&+&H.c., 
\label{eq:HtR} 
\end{eqnarray}
%----------------------------------------------------------------------
and 
%----------------------------------------------------------------------
\begin{eqnarray}
{\cal H}_V^{\rm{R}}&=&V_{{\rm{abNN}}} \sum\limits_{<ij>m} {n_{mi} n_{mj} }
 + V_{{\rm{cNN}}} \sum\limits_{(ij)} {n_{ui} n_{lj} } \nonumber\\
&+&V_{{\rm{cNNN}}} \sum\limits_{[ij]} {n_{ui} n_{lj} }. 
\label{eq:HVR} 
\end{eqnarray}
%----------------------------------------------------------------------
A schematic view of the lattice structure and the interactions are shown in Fig.~\ref{fig:mod2}. 
We consider the long-range Coulomb interactions between the in-plane NN sites ($V_{\rm abNN}$), those between the inter-plane NN sites ($V_{\rm cNN}$), and those between the inter-plane next-nearest-neighbor (NNN) sites ($V_{\rm cNNN}$). 
Symbols $\sum\nolimits_{<ij>}$, $\sum\nolimits_{(ij)}$ and 
$\sum\nolimits_{[ij]}$ represent summations for the pairs corresponding to $V_{\rm abNN}$, $V_{\rm cNN}$ and $V_{\rm cNNN}$, respectively. 
When the $1/r$-type Coulomb interaction is assumed, 
we have $V_{\rm cNN}/V_{\rm abNN}=1.2$ and $V_{\rm cNNN}/V_{\rm abNN}=0.77$ 
for the crystal structure in $R$Fe$_2$O$_4$. \cite{Kimizuka} 
As for the electron transfer term ${\cal H}_t^R$, 
the transfer integral between the in-plane NN sites ($t_{\rm ab}$), 
and that between the inter-plane NN sites ($t_{\rm c}$) are considered. 

%**********************************************************************
\begin{figure}[t]
%\vspace{-0.2cm}
\begin{center}
\includegraphics[width=7.5cm,height=8.5cm]{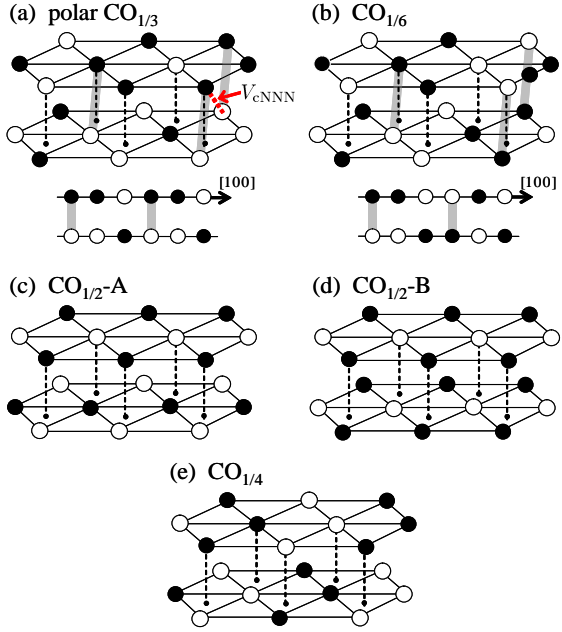}
\end{center}
\vspace{-0.4cm}
\caption{(Color online)
Schematic CO structures considered in ${\cal H}_{Vt}^{\rm R}$: 
(a) polar CO$_{1/3}$, (b) CO$_{1/6}$, 
(c) CO$_{1/2}$-A, (d) CO$_{1/2}$-B and (e) CO$_{1/4}$. 
The insets of (c) and (d) are side views along the [120] direction. 
}
\label{fig:CO2}
%\vspace{-0.5cm}
\end{figure}
%**********************************************************************

This model is analyzed by the VMC method. 
In the many-body correlation term ${\cal P}$, we consider the variational parameters $v_{\rm abNN}$, $v_{\rm cNN}$ and $v_{\rm cNNN}$ 
which correspond to the interactions $V_{\rm abNN}$, $V_{\rm cNN}$ and $V_{\rm cNNN}$, respectively, instead of Eq.~(\ref{eq:Pv}). 
In the one-body term of the wave function $\Phi$, we adopt the following CO structures: 
(i) a three-fold CO along the [100] direction termed CO$_{1/3}$, 
(ii) a six-fold CO along the [100] direction termed CO$_{1/6}$, 
(iii) two types of two-fold CO's along [100], i.e. (iii) CO$_{1/2}$-A 
and (iv) CO$_{1/2}$-B, and (v) a four-fold CO along [100] termed CO$_{1/4}$. 
These are schematically shown in Fig.~\ref{fig:CO2}. 
We define that the CO$_{1/3}$ structure in this realistic lattice has a polar CO configuration which is schematically given as  
$\cdots \bullet \circ \bullet \bullet \circ \bullet \cdots$ 
($\cdots \circ \bullet \circ \circ \bullet \circ \cdots$) along $[100]$ in the upper (lower) plane. 
Symbols $\circ$ and $\bullet$ are for the charge densities of $\langle n_{m i}-1/2 \rangle=\Delta$ 
and $-\Delta$, respectively, where $\Delta$ is the variational parameter in the one-body part of the wave function. 
When this CO is fully ordered, 
the electric polarization $p$ in Eq.~(\ref{eq:Ps}) is one.
The CO configuration along $[100]$ in the non-polar CO$_{1/6}$ structure is schematically given as 
$\cdots \bullet \circ \bullet \bullet \circ \circ \cdots$ 
($\cdots \circ \bullet \circ \circ \bullet \bullet \cdots$ ) 
in the upper (lower) plane. 
In each CO's, the one-body part of the wave function is obtained from the HF Hamiltonian. 
For example, the HF Hamiltonian for the polar CO$_{1/3}$ is given by  
%----------------------------------------------------------------------
\begin{eqnarray}
{\cal H}^{R\ 1/3}&=&\sum\limits_{{\bf{k}}m} {{\bf{\phi}}_{{\bf{k}}m}^\dag 
h_{{\bf{k}}m}^{{\rm R}} {\bf{\phi}}_{{\bf{k}}m} } \nonumber\\
&-&t_{\rm c} \sum\limits_{\bf{k}} {({\bf{\phi}}_{{\bf{k}}u}^\dag 
\xi_{\bf{k}} {\bf{\phi}}_{{\bf{k}}l} + H.c. } ), 
%{\bf{\phi}}_{{\bf{k}}l}^\dag C_{\bf{k}}^\dag  {\bf{\phi}}_{{\bf{k}}u} )} 
\label{eq:HMFR} 
\end{eqnarray}
%----------------------------------------------------------------------
with 
%%----------------------------------------------------------------------
%\begin{eqnarray}
%{\bf{t}}_{{\bf{k}}m} = \left( {\begin{array}{*{20}c}
%   {c_{{\bf{k}}m}^A }  \\
%   {c_{{\bf{k}}m}^B }  \\
%   {c_{{\bf{k}}m}^C }  \\
%\end{array}} \right), 
%\label{eq:tR} 
%\end{eqnarray}
%%----------------------------------------------------------------------
%----------------------------------------------------------------------
\begin{eqnarray}
h_{{\bf{k}}u}^{{\rm R}} = \left( {\begin{array}{*{20}c}
   {W_{1}^{\rm R}} & T_{\bf{k}}^{{\rm{R}}} & {T_{\bf{k}}^{{\rm{R}}*}}  \\
   {T_{\bf{k}}^{{\rm{R}}*}} & W_{1}^{\rm R} & T_{\bf{k}}^{{\rm{R}}}  \\
   T_{\bf{k}}^{{\rm{R}}} & {T_{\bf{k}}^{{\rm{R}}*}} & { - W_{2}^{\rm R}}  \\
\end{array}} \right), 
\label{eq:BR} 
\end{eqnarray} 
%----------------------------------------------------------------------
%----------------------------------------------------------------------
\begin{eqnarray}
h_{{\bf{k}}l}^{{\rm R}} = \left( {\begin{array}{*{20}c}
   { W_{2}^{\rm R} } & T_{\bf{k}}^{{\rm{R}}} & {T_{\bf{k}}^{{\rm{R}}*}}  \\
   {T_{\bf{k}}^{{\rm{R}}*}} & { - W_{1}^{\rm R}} & T_{\bf{k}}^{{\rm{R}}}  \\
   T_{\bf{k}}^{{\rm{R}}} & {T_{\bf{k}}^{{\rm{R}}*}} & { - W_{1}^{\rm R} }  \\
\end{array}} \right), 
\label{eq:BR2} 
\end{eqnarray} 
%----------------------------------------------------------------------
%----------------------------------------------------------------------
\begin{eqnarray}
\xi_{\bf{k}}  = \left( {\begin{array}{*{20}c}
   {e^{i\frac{{k_1  + 2k_2 }}{3}} } & {e^{i\frac{{2k_1  + k_2 }}{3}} } & {e^{i\frac{{k_1  - k_2 }}{3}} }  \\
   {e^{i\frac{{k_1  - k_2 }}{3}} } & {e^{i\frac{{k_1  + 2k_2 }}{3}} } & {e^{i\frac{{2k_1  + k_2 }}{3}} }  \\
   {e^{i\frac{{2k_1  + k_2 }}{3}} } & {e^{i\frac{{k_1  - k_2 }}{3}} } & {e^{i\frac{{k_1  + 2k_2 }}{3}} }  \\
\end{array}} \right), 
\label{eq:CR} 
\end{eqnarray} 
%----------------------------------------------------------------------
where 
%----------------------------------------------------------------------
\begin{eqnarray}
T_{\bf{k}}^{{\rm{R}}} = - t_{{\rm ab}}(e^{ik_1} + e^{ik_2} 
+ e^{-i(k_1 + k_2)}), 
\label{eq:TR}
\end{eqnarray}
%----------------------------------------------------------------------
%----------------------------------------------------------------------
\begin{eqnarray}
W_{1}^{{\rm{R}}} = - \Delta (V_{{\rm{cNN}}} + V_{{\rm{cNNN}}}) , 
\label{eq:VM1} 
\end{eqnarray} 
%----------------------------------------------------------------------
and 
%----------------------------------------------------------------------
\begin{eqnarray}
W_{2}^{{\rm{R}}} = \Delta (6V_{{\rm{abNN}}} - 
V_{{\rm{cNN}}} - V_{{\rm{cNNN}}}). 
\label{eq:VM2} 
\end{eqnarray} 
%----------------------------------------------------------------------
The summation $\sum\nolimits_{\bf k}$ runs over the Brillouin zone 
for the CO structure with the momentum $(2\pi/3,2\pi/3)$. 
The charge densities in sublattices A and B in the upper plane and 
that in C in the lower plane are $\Delta(>0)$, and those in other sublattices are $-\Delta$. 
The one-body parts of the wave functions in other CO's are also obtained in the same way. 

%**********************************************************************
\begin{figure}[t]
%\vspace{-0.2cm}
\begin{center}
\includegraphics[width=8.2cm,height=6.0cm]{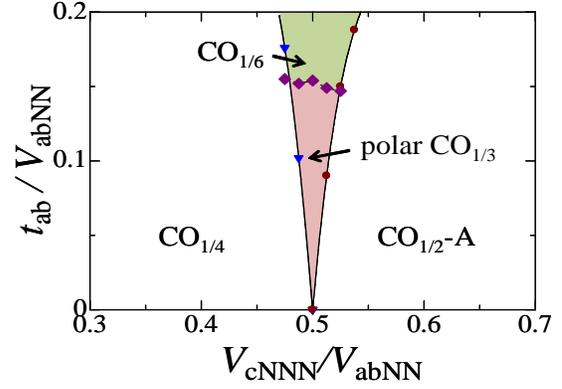}
\end{center}
\vspace{-0.8cm}
\caption{(Color online)
Phase diagram in ${\cal H}_{Vt}^{\rm R}$. 
Parameters are chosen to be $t_{\rm c}/t_{\rm ab} = 1$ and $V_{\rm cNN}/V_{\rm abNN} = 1$. 
}
\label{fig:ph}
%\vspace{-0.5cm}
\end{figure}
%**********************************************************************

%**********************************************************************
\begin{figure}[t]
%\vspace{-0.2cm}
\begin{center}
\includegraphics[width=7.3cm,height=7.5cm]{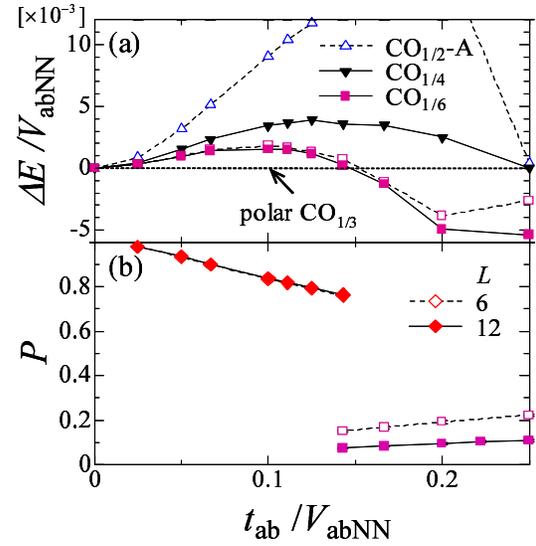}
\end{center}
\vspace{-0.5cm}
\caption{(Color online) 
(a) Differences between the energies in several CO's and that in the CO$_{1/3}$ 
obtained in ${\cal H}_{Vt}^{\rm R}$. 
(b) Polarization correlation $P$. 
Parameters are chosen to be $t_{\rm ab}=t_{\rm c}$ and $V_{\rm abNN}=V_{\rm cNN}=2V_{\rm cNNN}$. 
Open and filled symbols are for the results in $L = 6$ and $12$, respectively.
}
\label{fig:eP}
%\vspace{-0.5cm}
\end{figure}
%**********************************************************************

In Fig.~\ref{fig:ph}, the phase diagram in the plane of $V_{\rm cNNN}/V_{\rm abNN}$ and $t_{\rm ab}/V_{\rm abNN}$ is presented. 
We chose the relations $V_{\rm cNN}=V_{\rm abNN}$ and $t_{\rm ab}=t_{\rm c}$, 
although the results introduced below are robust qualitatively around these parameter.  
This phase diagram is obtained by the energy and the charge correlation function introduced in Eq.~(\ref{eq:Nq}). 
A point of $V_{\rm cNNN}/V_{\rm abNN}=0.5$ at $t_{\rm ab}/V_{\rm abNN}=0$ is a fully frustrated point which corresponds to $V'/V=1$ and $t/V=0$ in Fig.~\ref{fig:simph}. 
In this point, the polar CO$_{1/3}$ is degenerate with CO$_{1/4}$ and CO$_{1/2}$-A. 
It is shown that the polar CO$_{1/3}$ is realized in finite parameter region, when the electron transfer is taken into account. 

We show the results in $V_{\rm cNNN}/V_{\rm abNN}=0.5$ in more detail.  
In Fig.~\ref{fig:eP}(a), we plot the energy differences of the several CO's from that in the CO$_{1/3}$ structure. 
The result for CO$_{1/2}$-B is much larger than the scale in the figure. 
The polar CO$_{1/3}$ phase competes mainly with the non-polar CO$_{1/6}$,  
and is stabilized in $t_{\rm ab}/V_{\rm abNN} \lsim 0.15$. 
The electric polarization correlation is presented in Fig.~\ref{fig:eP}(b). 
The results in the CO$_{1/3}$ phase are independent of the system size $L$, and show almost its maximum value. 
On the other hand, the value of $P$ in the CO$_{1/6}$ phase are much smaller than those in the CO$_{1/3}$ phase and decrease with increasing of the system size. 

\begin{figure}[t]
%\vspace{-0.2cm}
\begin{center}
\includegraphics[width=7.3cm,height=3.5cm]{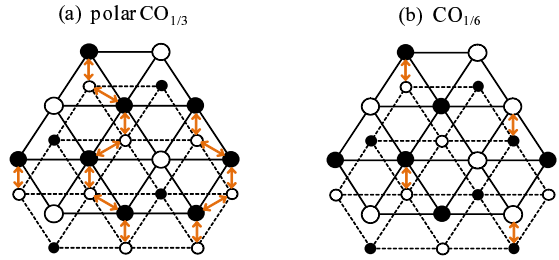}
\end{center}
\vspace{-0.4cm}
\caption{(Color online)
Inter-layer charge fluctuation processes by 
$t_{\rm c}$ in the polar CO$_{1/3}$ phase (a) and those in the CO$_{1/6}$ phase (b). 
The bold arrows represent the charge fluctuation 
where the energy reduction is  
of the order of $t_{\rm c}^2/V_{\rm cNNN}$. 
Large (small) circles represent the sites in the upper (lower) layer. 
}
\label{fig:comp}
%\vspace{-0.5cm}
\end{figure}

A stability of the polar CO$_{1/3}$ is attributed to a combination of the inter-layer 
charge transfer, $t_{\rm c}$, and the long-range Coulomb interaction, $V_{\rm cNNN}$, in this realistic crystal structure as follows. 
Let us focus on the bonds represented by the gray bars in the Fig.~\ref{fig:CO2}(a). 
These bonds connect the black-and-white circles and are responsible for the electric dipole moments. 
The inter-layer Coulomb interaction $V_{\rm cNNN}$ causes the interaction along the $[120]$ direction between the bonds, and induces the ferroelectric interaction between the dipole moments along this direction. 
Although the dipole moments are induced in the local bonds, 
there is still competition between the polar CO$_{1/3}$ 
and the antiferro-electric CO$_{1/6}$ 
which are in the different configurations of the dipole moments along [100], 
as shown in Figs.~\ref{fig:CO2}(a) and (b). 
This degeneracy is lifted by $t_{\rm c}$ as follows.  
In Fig.~\ref{fig:comp}, we compare the inter-layer charge fluctuation process by $t_{\rm c}$ in CO$_{1/3}$ with that in CO$_{1/6}$.
The energy reduction due to this type of the charge fluctuation 
is of the order of $t_{\rm c}^2/V_{\rm cNNN}$;  
this fluctuation easily occurs, because the intermediate-state energy $V_{\rm cNNN}$ is the smallest Coulomb interaction in the model of ${\cal H}_{Vt}^{\rm R}$. 
In the CO$_{1/3}$ structure, the charge fluctuation occurs at the sites represented by the black (white) circles connected by arrows in the upper (lower) layer in Fig.~\ref{fig:comp}(a). 
On the other hand, in the case of the CO$_{1/6}$ structure, the number of sites where this type of charge fluctuation occurs are half of that in CO$_{1/3}$, as shown in Fig.~\ref{fig:comp}(b). In other CO's, i.e. CO$_{1/2}$'s and CO$_{1/4}$, this fluctuation process is prohibited. 
%That is, stability of the polar CO$_{1/3}$ phase is attributed to the inter-layer charge fluctuation.  

%Consider the charge fluctuation at the sites represented by 
%the black (white) circles in the upper (lower) layer 
%in Fig.~\ref{fig:CO2}(a). 
%In the inter-layer charge fluctuation by $t_{\rm c}$, 
%the energy reduction is of the order of $t_{\rm c}^2/V_{\rm cNNN}$ 
%where the intermediate-state energy $V_{\rm cNNN}$ is the smallest 
%Coulomb interaction in the present model of ${\cal H}_{Vt}^{\rm R}$. 
%Therefore, this kind of fluctuation easily occurs and stabilizes 
%the polar CO$_{1/3}$. 
%The kind of charge fluctuation also appears 
%in the CO$_{1/6}$, but the number of sites at which the 
%fluctuation occur is half of that in the polar CO$_{1/3}$. 
%Furthermore, in the other CO's, i.e. CO$_{1/2}$'s and CO$_{1/4}$, 
%this fluctuation process is prohibited. 
%Thus, the polar CO$_{1/3}$ is realized as the most stable state 
%in the present model. 

%**********************************************************************
%**********************************************************************

In order to support this scenario, we calculate the energies by the second-order perturbation in terms of $t_{\rm ab}$ and $t_{\rm c}$. 
A general form of the energies in the CO's, shown in Fig.~\ref{fig:eP}(a), are given by 
$E/V_{{\rm{abNN}}} = 2 - \alpha (t_{{\rm{ab}}} /V_{{\rm{abNN}}} )^2
 - \beta (t_{\rm{c}} /V_{{\rm{abNN}}} )^2$ 
where $\alpha$ and $\beta$ are positive numerical constants. 
The second and third terms imply energy gains from the intra-layer and inter-layer charge 
fluctuations, respectively. 
We have $(\alpha, \beta)=$(2.00,2.76) in polar CO$_{1/3}$, (2.60,1.83) in CO$_{1/6}$, 
(2.67,0.67) in CO$_{1/2}$-A and (2.67,1.33) in CO$_{1/4}$. 
That is, in the polar CO$_{1/3}$, $\alpha$ is smallest, 
while $\beta$ is largest among them. 
This indicates that the stabilization of the polar CO$_{1/3}$ phase  
is mainly caused by the inter-layer charge fluctuation due to $t_{\rm c}$.
 
%On the other hand, appearance of the nonpolar CO$_{1/6}$ for $t_{\rm ab}/V_{\rm abNN} \gsim 0.15$, as seen in Fig.~\ref{fig:ph}, can't be understood from the present result. 
%We suppose that higher-order terms of $t_{\rm ab}$ are responsible for the stabilization of CO$_{1/6}$. 

%as a function of $t_{\rm ab}/V_{\rm abNN}$, 
%where $t_{\rm ab}=t_{\rm c}$ and $V_{\rm abNN}=V_{\rm cNN}=2V_{\rm cNNN}$
%are assumed. 
%By introducing $t_{\rm ab}$, the energy for the polar CO$_{1/3}$ becomes 
%lowest in the range of $t_{\rm ab}>0$. 
%The formula for the energies is given as 
%$E/V_{{\rm{abNN}}} = 2 - \alpha (t_{{\rm{ab}}} /V_{{\rm{abNN}}} )^2
% - \beta (t_{\rm{c}} /V_{{\rm{abNN}}} )^2$, 
%where $\alpha$ and $\beta$ are positive constant numbers. 
%The second term implies an energy gain from the in-plane charge 
%fluctuation, whereas the third term is the energy gain from the 
%inter-layer fluctuation. 
%In the polar CO$_{1/3}$, the value of $\alpha$ is smallest 
%among all the CO's ($\alpha=2.00$), while $\beta$ is largest 
%among them ($\beta=2.76$). 
%This indicates that the stabilization of the polar CO$_{1/3}$ 
%is mostly due to the inter-layer charge fluctuation by $t_{\rm c}$, 
%as we considered above. 
%On the other hand, appearance of the nonpolar CO$_{1/6}$ for 
%$t_{\rm ab}/V_{\rm abNN} \gsim 0.15$, 
%as seen in Fig.~\ref{fig:mod2ph}(b), can't be understood from 
%the present result in Fig.~\ref{fig:per}. 
%We suppose that higher-order terms of $t_{\rm ab}$ are responsible for 
%the stabilization of CO$_{1/6}$. }

%%%%%%%%%%%%%%%%%%%%%%%%%%%%%%%%%%%%%%%%%%
\section{Summary and Discussion 
\label{sec:summary}}
%%%%%%%%%%%%%%%%%%%%%%%%%%%%%%%%%%%%%%%%%%

In this section, we discuss the relation between the present results and our previous results in Ref.~\cite{Naka}, and their implications for the layered iron oxides. 
In our previous examinations, we study the dielectric properties in a paired-triangular lattice; 
the long-range Coulomb interactions are introduced between the electronic charges which are treated as classical variables. 
The model Hamiltonian was analyzed by the mean-field method and the classical Monte-Carlo simulation in finite size clusters. 
It was shown that the three-fold charge ordered state, i.e. CO$_{1/3}$, does not appear at zero temperature except for the fully frustrated point, corresponding to $V_{\rm cNNN}/V_{\rm abNN}=0.6$ in Fig.~7 in Ref.~\cite{Naka} and $V_{\rm cNNN}/V_{\rm abNN}=0.5$ in Fig.~\ref{fig:ph}, 
and is stabilized in finite temperatures due to thermal fluctuation. 
The obtained V-shape phase diagram in the temperature and $V_{\rm cNNN}/V_{\rm abNN}$ plane (see Fig.~7 in Ref.~\cite{Naka}) is similar to that in Fig.~\ref{fig:ph} where the electron transfer integral corresponds to the temperature. 
This comparison suggests that the thermal and quantum fluctuations play similar roles for stabilization of the polar three-fold charge ordered state.  
Naively speaking, the CO$_{1/3}$ state is not a robust charge ordered state, since, as shown in Fig.~\ref{fig:CO2}(a), 
the intra-layer NN Coulomb interactions at black (white) circles in the upper (lower) plane are canceled out. Therefore, the thermal motion and the quantum virtual motion of charges at these sites easily occur, and these contribute to the entropy gain and the transfer energy gain in the classical and quantum cases, respectively. 
This is the so-called order-by-fluctuation process. 

One notable difference between the two types of the fluctuation is that the polarization correlation $P$ defined in Eq.~(\ref{eq:Pl}) is weaken with increasing system size in the finite temperature classical calculation, while it is robust with increasing $L$ in the present quantum calculation at $T=0$. 
In the classical Monte Carlo simulation in finite temperatures, a large number of non-polar structure coexists with the polar CO$_{1/3}$ structure. 
As a result, the thermal expectation value of the polarization correlation tends to decrease with the system size. 
On the other hand, in the present VMC calculation where the long-rage charge ordered states are assumed in the variational wave function, 
the fully polarized charge ordered state is energetically stabilized due to the inter-layer electron transfer and the realistic crystal structure, as explained above. 

Through these two theoretical examinations, i.e. the classical Mote-Carlo simulation in finite temperature and the VMC simulation at $T=0$, we conclude that charge fluctuation plays essential roles in the dielectric properties in layered iron oxides. 
These theoretical results have implications for the experimental results in layered iron oxides. The charge fluctuation between Fe$^{2+}$ and Fe$^{3+}$ has been observed in the optical absorption experiments and the M$\rm \ddot o$ssbauer spectroscopy measurements.~\cite{Xu,Nakamura} It was suggested in the temperature dependence of the optical absorption spectroscopy that the charge fluctuation remains to be large even far below the charge ordering temperature. This is consistent with our scenario that the thermal/quantum charge fluctuation stabilizes the polar charge order. 
The diffusive nature in the dielectric constant observed near the charge ordering temperature is also related to the remarkable charge fluctuation predicted in the CO$_{1/3}$ state. 
Another relation of the present results to the layered iron oxides is the several different charge ordered states observed by changing the rare-earth ion in $R$Fe$_2$O$_4$. 
In LuFe$_2$O$_4$, the three-fold charge order associated is stabilized below 320K. On the other hand, in YFe$_2$O$_4$, a sequential phase transition occurs as (the charge disordered state) $\rightarrow$ (the polar three-fold CO) $\rightarrow$ (the non-polar CO's with different periodicities) by decreasing temperature.~\cite{ikeda03} That is, the polar CO phase only appears in a middle range of temperature. 
This result is consistent with our results of Fig.~\ref{fig:ph} and Fig.~7 in our previous paper~\cite{Naka}, when we speculate that a change of the $R$ ion corresponds to a change of the Coulomb interaction parameter $V_{\rm cNNN}/V_{\rm abNN}$ and/or the electron transfer intensity. 

In summary, we examine roles of the quantum fluctuation, i.e. the electron transfer effect, in the electronic ferroelectricity. 
In particular, we focus on the dielectric properties in a paired triangular lattice, motivated by the multiferroic layered iron oxides. 
The variational Monte-Carlo simulation is applied to the three types of the extended $V-t$ models. 
It is found that the quantum transfer between the triangular lattice layers tends to promote the three-fold polar charge ordered state. 
Both the spin degree of freedom and the realistic crystal structure in the layered iron oxides reinforce the electric polarization. 
Present roles of the quantum fluctuation on the ferroelectric transition are in highly contrast to the conventional manner of the quantum dielectric fluctuation in the hydrogen-bond type ferroelectricity and the quantum paraelectric oxides. 

Authors would like to thank S.~Mori, N.~Ikeda, H.~Takashima, 
M.~Naka and J.~Nasu for their valuable discussions. 
This work was supported by JSPS KAKENHI, TOKUTEI from MEXT, 
and Grand challenges in next-generation integrated nanoscience.

\noindent

%*****************************************************************************

%*****************************************************************************

\end{document}